\documentclass[a4paper,twoside,12pt]{article}
\input{obsjkt.sty}

\newcommand{\Msun}{\ensuremath{\,{\rm M}_\odot}}                  
\newcommand{\Rsun}{\ensuremath{\,{\rm R}_\odot}}                  
\newcommand{\rhosun}{\ensuremath{\,\rho_\odot}}                   
\newcommand{\Teff}{\ensuremath{T_{\rm eff}}}                      
\newcommand{\FeH}{\ensuremath{\rm [Fe/H]}}                        
\newcommand{\degr}{\ensuremath{^\circ}}                           
\renewcommand{\kms}{\,km\,s$^{-1}$}                               
\newcommand{\as}{\ensuremath{^{\prime\prime}}}                    
\newcommand{\etal}{\textit{et al.}}                               

\newcommand{\kepler}{\textit{Kepler}}
\newcommand{\tess}{\textit{TESS}}
\newcommand{\hip}{\textit{Hipparcos}}
\newcommand{\gaia}{\textit{Gaia}}

\newcommand{\Msunnom}{\hbox{$\mathcal{M}^{\rm N}_\odot$}}
\newcommand{\Rsunnom}{\hbox{$\mathcal{R}^{\rm N}_\odot$}}
\newcommand{\Lsunnom}{\hbox{$\mathcal{L}^{\rm N}_\odot$}}

\begin{document} 

\OBSheader{Rediscussion of eclipsing binaries: V455 Aur}{J.\ Southworth}{2021 Aug}

\OBStitle{Rediscussion of eclipsing binaries. Paper V. \\ The triple system V455\,Aurigae}

\OBSauth{John Southworth}

\OBSinstone{Astrophysics Group, Keele University, Staffordshire, ST5 5BG, UK}

\OBSabstract{V455\,Aur is a detached eclipsing binary containing two F-stars in a 3.15\,d orbit with a small eccentricity. Its eclipses were discovered in data from the \hip\ satellite and a spectroscopic orbit was obtained by Griffin \cite{Griffin01obs,Griffin13obs}. Griffin found a long-term variation of the systemic velocity of the eclipsing system due to a third body in a highly eccentric orbit ($e=0.73$) with a period of 4200\,d. We have used these data, the light curve of V455\,Aur from the \tess\ satellite, and the \gaia\ EDR3 parallax to determine the physical properties of the components of the system to high precision. We find the eclipsing stars to have masses of $1.289 \pm 0.006$\Msun\ and $1.232 \pm 0.005$\Msun, radii of $1.389 \pm 0.011$\Rsun\ and $1.318 \pm 0.014$\Rsun\ and effective temperatures of $6500 \pm 200$ and $6400 \pm 200$\,K. Light from the tertiary component is directly detected for the first time, in the form of a third light of $\ell_3 = 0.028 \pm 0.002$ in the solution of the \tess\ light curve. From this $\ell_3$, theoretical spectra and empirical calibrations we estimate the star to have a mass of $0.72 \pm 0.05$\Msun, a radius of $0.74 \pm 0.05$\Rsun\ and a temperature of $4300 \pm 300$~K. The inclination of the outer orbit is $53 \pm 3\degr$, so the two orbits in the system are not coplanar. We show that a measured spectroscopic light ratio of the two eclipsing stars could lower the uncertainties in radius from 1\% to 0.25\%. A detailed spectroscopic analysis could also yield precise temperatures and chemical abundances of the system, thus making V455\,Aur one of the most precisely measured eclipsing systems known.}


\section*{Introduction}

The study of detached eclipsing binaries (dEBs) is a mature area of research \cite{Russell14nat,Popper67araa,Popper80araa,Andersen91aarv,Torres++10aarv} that provides foundational measurements of the physical properties of normal stars against which our theoretical understanding can be compared, verified and improved \cite{Andersen++90apj,Pols+97mn,Spada+13apj,ClaretTorres18apj,Tkachenko+20aa}. Many dEBs have long observational histories, stretching in some cases to over a century \cite{Stebbins11apj,Stebbins21apj}. For these objects, and many others, the ongoing NASA Transiting Exoplanet Survey Satellite \cite{Ricker+15jatis} (\tess) mission is providing light curves of previously unobtainable quality. Whilst the \tess\ data are relatively poor in comparison to the NASA \kepler\ satellite \cite{Borucki16rpph}, they cover a much larger sky area and thus capture information on many more dEBs than \kepler\ did. \tess\ light curves have been used to provide high-quality measurements of the radii of the components of dEBs \cite{Maxted+20mn,Graczyk+21aa}, discover pulsations in well-known dEBs \cite{Me+20mn,LeeHong21aj,Me++21mn,Budding+21mn,Lee+21aj}, obtain eclipse timings \cite{Vonessen+20xxx}, study apsidal motion \cite{Baroch+21aa}, discover multiply-eclipsing binaries \cite{Borkovits+20mn,Borkovits+21mn} and investigate heartbeat stars \cite{Kolaczek+21aa}.

Within this context we have begun a project to utilise the \tess\ database of time-series photometry to revise and improve the measured physical properties of known dEBs \cite{Me20obs,Me21obs1,Me21obs2,Me21obs3}. The principle aim is to curate the Detached Eclipsing Binary Catalogue\footnote{\texttt{https://www.astro.keele.ac.uk/jkt/debcat/}} (DEBCat) of dEBs with masses and radii measured to precisions of 2\% or better \cite{Me15debcat}, by reanalysing existing systems or by adding new dEBs for which good radial velocity (RV) curves already exist.


\begin{table}[t]
\caption{\em Basic information on V455\,Aur \label{tab:info}}
\centering
\begin{tabular}{lll}
{\em Property}                      & {\em Value}           & {\em Reference}                   \\[3pt]
Henry Draper designation            & HD 45191              & \cite{CannonPickering18anhar2}    \\
\textit{Hipparcos} designation      & HIP 30878             & \cite{Hipparcos97}                \\
\textit{Tycho} designation          & TYC 3388-1017-1       & \cite{Hog+00aa}                   \\
\textit{Gaia} EDR3 designation      & 992435451683384320    & \cite{Gaia20aa}                   \\
\textit{Gaia} EDR3 parallax         & $12.874\pm 0.052$ mas & \cite{Gaia20aa}                   \\
$B$ magnitude                       & $7.71  \pm 0.01 $     & \cite{Hog+00aa}                   \\
$V$ magnitude                       & $7.28  \pm 0.01 $     & \cite{Hog+00aa}                   \\
$J$ magnitude                       & $6.353 \pm 0.019$     & \cite{Cutri+03book}               \\
$H$ magnitude                       & $6.132 \pm 0.038$     & \cite{Cutri+03book}               \\
$K_s$ magnitude                     & $6.082 \pm 0.024$     & \cite{Cutri+03book}               \\
Spectral type                       & F5\,V + F6\,V         & This work                         \\[10pt]
\end{tabular}
\end{table}

\section*{V455\,Aurigae}

\begin{figure}[t] \centering \includegraphics[width=\textwidth]{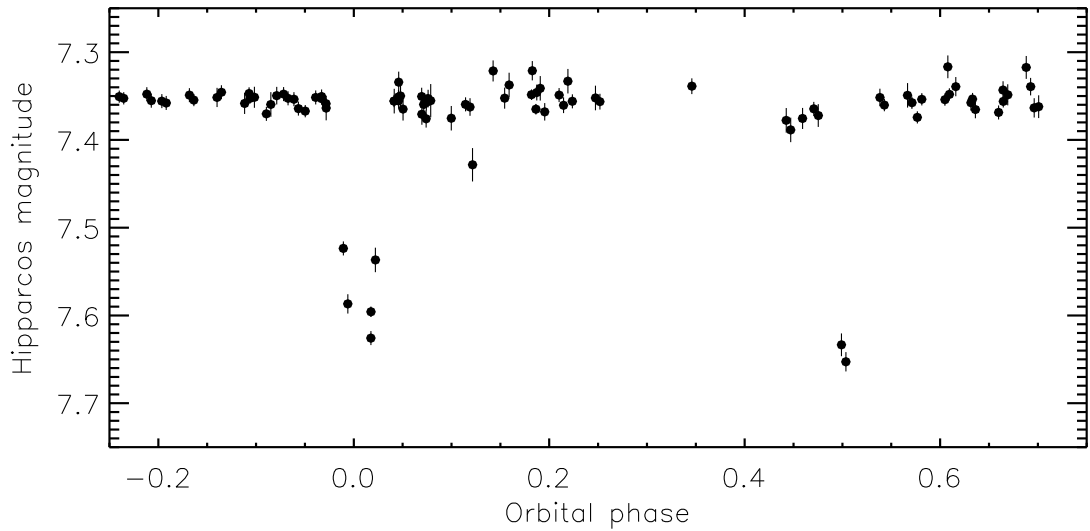} \\
\caption{\label{fig:hip} The \hip\ light curve of V455\,Aur plotted as a function 
of orbital phase using the ephemeris determined in the current work.} \end{figure}

In this work we present the first measurements of the masses and radii of the components of V455\,Aur (Table\,\ref{tab:info}). This object was discovered to be eclipsing using \hip\ data \cite{Hipparcos97} and given its variable-star designation shortly afterwards \cite{Kazarovets+99ibvs}. Fig.\,\ref{fig:hip} shows a plot of the \hip\ epoch photometry.

Griffin \cite{Griffin01obs} presented the first spectroscopic orbit of the system, based on photoelectric observations from the Cambridge \textit{CORAVEL}. A total of 42 RVs were measured for each star. The relatively early spectral type for this instrument proved not to be a problem due to the brightness of the system, although the dips in the cross-correlation functions (CCFs) were only approximately 4\% deep (his fig.\,4). The resulting spectroscopic orbits provided the first orbital period determination for this system plus measurement of the minimum masses ($M_{\rm A,B} \sin^3i$ where $i$ is the orbital inclination) to better than 1\% precision. Griffin \cite{Griffin01obs} noticed a change in the systemic velocity between the two observing seasons during which he obtained data, and surmised the presence of a third star on a wider orbit. He also suggested that the spectral type of F2 from the Henry Draper Catalogue \cite{CannonPickering18anhar2} was too early, and that a type of approximately F6\,V + F7\,V better reflected the properties of the system.

Griffin \cite{Griffin13obs} revisited the system with additional RV measurements in order to establish the properties of the outer orbit. From 123 RVs of each of the two eclipsing stars he obtained minimum masses to 0.5\% precision and confirmed the small but significant orbital eccentricity of the inner orbit. The outer orbit was found to have a period of $4205 \pm 17$\,d, an amplitude of $5.05 \pm 0.20$\kms\ and a high eccentricity of $0.726 \pm 0.017$. The CCF dips of the two stars have a ratio of 1.21 and this corresponds to a magnitude difference of 0.24\,mag in the $V$ band; this is of course a measure of the relative strengths of the spectral lines of the two stars, not their continuum brightnesses. The mean projected rotational velocities were found to be $17.98 \pm 0.17$\kms\ and $18.77 \pm 0.29$\kms. Griffin \cite{Griffin13obs} constrained the mass of the tertiary component to be more than 0.5\Msun\ based on the mass function of the outer orbit, but of spectral type K or later due to the absence of a signal in the CCF.

There are only two other studies that provide relevant information on V455\,Aur. Casagrande \etal\ \cite{Casagrande+11aa} obtained an effective temperature of $\Teff = 6405 \pm 80$\,K and a metallicity of $\FeH = -0.25$ for the system using calibrations based on Str\"omgren photometry. Binarity was not accounted for in the determination of these values, so they should be treated with caution. The \Teff\ listed in the \tess\ Input Catalogue (TIC) version 8.0 (Stassun \etal\ \cite{Stassun+19aj}) is $6406 \pm 142$\,K. One time of minimum light has been observed by the BAV \cite{Hubscher15ibvs}.


\section*{Observational material}

\begin{figure}[t] \centering \includegraphics[width=\textwidth]{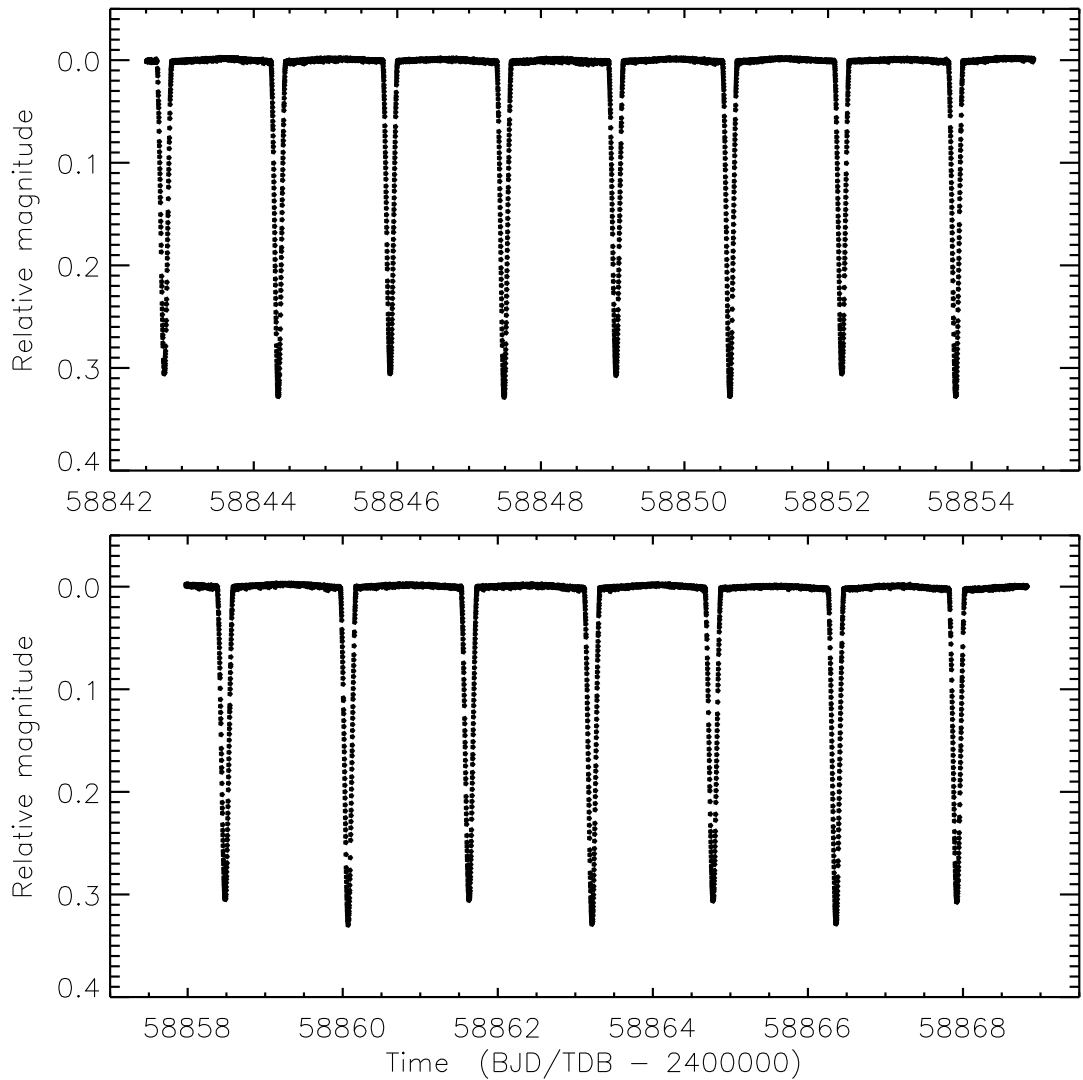} \\
\caption{\label{fig:time} \tess\ Sector 20 short-cadence photometry of V455\,Aur.
The two panels show the data before and after the mid-sector pause for download
of the data to Earth \cite{Ricker+15jatis}.} \end{figure}


The data used in this work comprise a light curve from camera 1 of the NASA \tess\ satellite \cite{Ricker+15jatis}, which observed V455\,Aur in Sector 20 (2019/12/24 to 2020/01/21). The light curve covers seven primary and eight secondary eclipses. These data were downloaded from the MAST archive\footnote{Mikulski Archive for Space Telescopes, \\ \texttt{https://mast.stsci.edu/portal/Mashup/Clients/Mast/Portal.html}} and converted to relative magnitude. We retained only those datapoints with the QUALITY flag equal to zero. The contamination ratio in the TIC\,v8.0 is 0.011 which indicates very little contaminating light in the \tess\ photometric aperture.

The \tess\ data were obtained in short cadence mode, with a sampling rate of 120\,s, and include 16\,412 datapoints (Fig.\,\ref{fig:time}). They are available, as usual, in two flavours: simple aperture photometry (SAP) and pre-search data conditioning (PDC) \cite{Jenkins+16spie}. As with previous papers in this series, we have chosen to adopt the SAP data for further analysis. This is because the eclipse shapes are indistinguishable between the two datasets but the PDC data contain jumps and slow undulations that have been introduced by the PDC correction process.


\section*{Light curve analysis}

\begin{figure}[t] \centering \includegraphics[width=\textwidth]{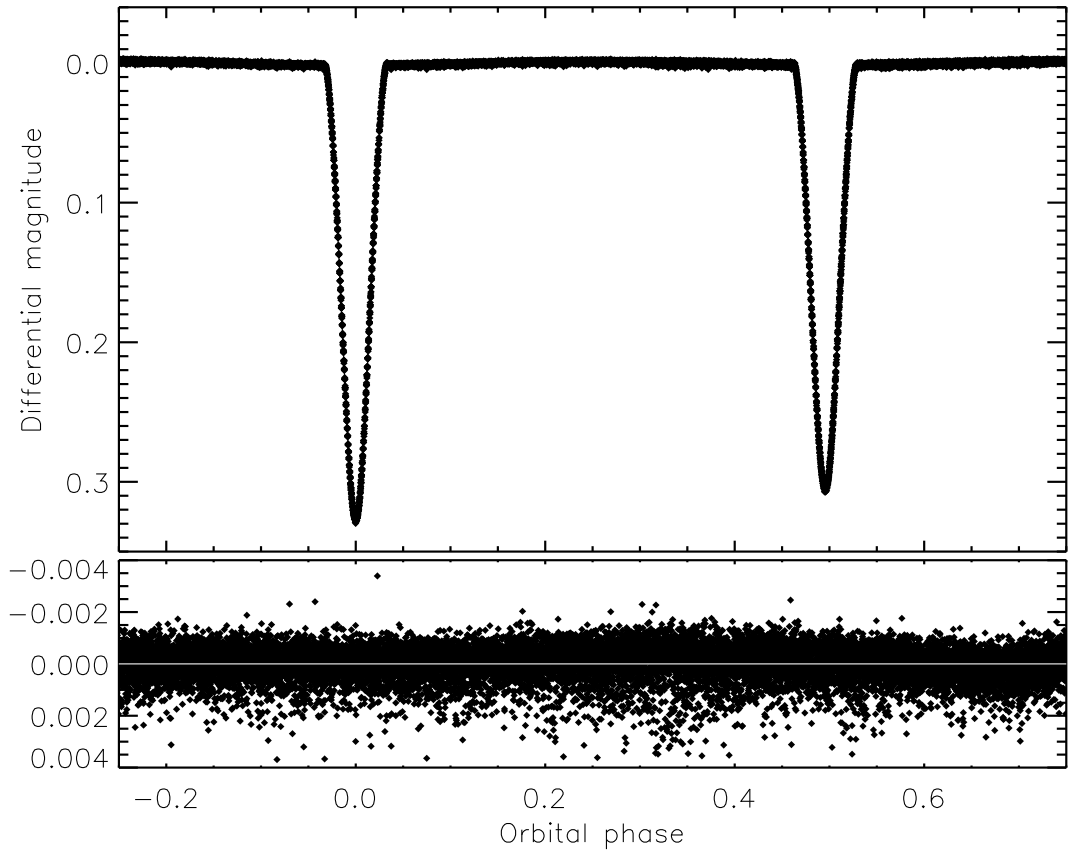} \\
\caption{\label{fig:tess} The \tess\ light curve of V455\,Aur (filled circles)
compared to the {\sc jktebop} fit (solid line, not visible in front of the data).
The lower panels show the residuals of the fit with the line of zero residual
overplotted in white for clarity.} \end{figure}

The components of V455\,Aur are small compared to their separation so we adopted version 41 of the {\sc jktebop}\footnote{\texttt{http://www.astro.keele.ac.uk/jkt/codes/jktebop.html}} code \cite{Me++04mn2,Me13aa} for our analysis of the \tess\ light curve. {\sc jktebop} is fast and flexible, and agrees well with more sophisticated codes when the stars are well-detached \cite{Maxted+20mn}. We label the star eclipsed at the primary (deeper) eclipse as star A, and its companion as star B; in this case star A is hotter, larger and more massive than star B.

We modelled the \tess\ data and did not consider any additional information such as RVs or times of minimum light. This is because the system has a third body which modifies the observed RVs and times of eclipse of the system, and measuring this change is not a goal of the current work. We fitted for the orbital period and a reference time of primary mid-eclipse. We also fitted for the fractional radii ($r_{\rm A} = \frac{R_{\rm A}}{a}$ and $r_{\rm B} = \frac{R_{\rm B}}{a}$ where $R_{\rm A}$ and $R_{\rm B}$ are the true radii of the stars, and $a$ is the semimajor axis of the relative orbit), parameterised as their sum ($r_{\rm A}+r_{\rm B}$) and ratio ($k = \frac{r_{\rm B}}{r_{\rm A}}$), the orbital inclination ($i$) and the central surface brightness ratio ($J$). As V455\,Aur is a triple system we also fitted for the third light ($\ell_3$) to account for the contribution of the fainter tertiary component; $\ell_3$ is defined as the fractional amount of the total light from the system at phase 0.25 that is not contributed by the two eclipsing stars.

The limb darkening was accounted for using the quadratic law \cite{Kopal50}, for which we took coefficients from Claret\cite{Claret18aa}. We fixed the quadratic coefficient to the theoretical value but fitted for the linear coefficient, requiring the coefficients to be the same for both stars due to their similarity in \Teff\ and surface gravity. Fixing one of the two coefficients is acceptable as they are strongly correlated \cite{Me++07aa,Me08mn} so any imprecision in one coefficient is easily accounted for by a similar change in the other.

A small but highly significant orbital eccentricity ($e$) was found by Griffin \cite{Griffin01obs,Griffin13obs} and we find that this is also required in order to get a good fit to the \tess\ data. We accounted for the eccentric orbit by fitting for the parameters $e\cos\omega$ and $e\sin\omega$, where $\omega$ is the argument of periastron. Finally, we applied polynomial fits to the overall brightness of the system to deal with any slow changes in brightness, likely of instrumental origin, during the \tess\ observations. We used two first-order polynomials, one for each half of the data (i.e.\ the two panels in Fig.\,\ref{fig:time}) and fitted the coefficients simultaneously with the other parameters in {\sc jktebop}.

\begin{table} \centering
\caption{\em \label{tab:lc} Parameters of the best {\sc jktebop} fit to the \tess\ light curve of V455\,Aur.
The uncertainties are 1$\sigma$. The same limb darkening coefficients were used for both stars.}
\begin{tabular}{lr@{\,$\pm$\,}l}
{\em Parameter}                           & \multicolumn{2}{c}{\em Value}    \\[3pt]
{\it Fitted parameters:} \\
Primary eclipse time (BJD/TDB)            & 2458853.778650  & 0.000008        \\
Orbital period (d)                        &      3.145777   & 0.000004        \\
Orbital inclination (\degr)               &      84.997     & 0.021           \\
Sum of the fractional radii               &       0.22018   & 0.00024         \\
Ratio of the radii                        &       0.949     & 0.017           \\
Central surface brightness ratio          &       0.9543    & 0.0011          \\
Third light                               &       0.0282    & 0.0022          \\
Linear limb darkening coefficient         &       0.221     & 0.012            \\
Quadratic limb darkening coefficient      & \multicolumn{2}{c}{~~~~~~~~~~0.22 (fixed)}  \\
$e\cos\omega$                             &    $-$0.006803  & 0.000007        \\
$e\sin\omega$                             &       0.00712   & 0.00037         \\
{\it Derived parameters:} \\
Fractional radius of star A               &       0.1130    & 0.0009          \\
Fractional radius of star B               &       0.1072    & 0.0011          \\
Orbital eccentricity                      &       0.00985   & 0.00027         \\
Argument of periastron (\degr)            &     133.7       & 1.5             \\
Light ratio                               &       0.859     & 0.031           \\
\end{tabular}
\end{table}

The best {\sc jktebop} fit to the \tess\ data is shown in Fig.\,\ref{fig:tess} and is very good. The residuals have a hint of a non-Gaussian distribution to them -- they scatter more to fainter than brighter magnitudes -- an  effect that we have often seen in \tess\ light curves. The parameters of the fit are given in Table\,\ref{tab:lc}. The secondary eclipse occurs at orbital phase 0.4967.


\section*{Determinacy of the photometric parameters}

V455\,Aur is a system where one might expect to find significant correlations between parameters, as it exhibits partial eclipses and third light. In addition, it has an eccentric orbit and $e\sin\omega$ is often correlated with the ratio of the radii and the orbital inclination. We therefore explored this in detail.

First, we determined the uncertainties in the photometric parameters using three approachs: the Monte Carlo and residual-permutation algorithms in \textsc{jktebop} \cite{Me++04mn,Me08mn} and by fitting subsections of the light curve in isolation. For the Monte Carlo and residual-permutation algorithms we calculated 10\,000 individual solutions. For the fitting-subsections approach we broke the \tess\ data into five sets, each containing three consecutive eclipses, and modelled each in the same way as for the whole dataset. As a result, each fitted and derived parameter had three estimates of its uncertainty, and we simply adopted the largest of the three possibilities (see Table\,\ref{tab:lc}). All three error estimates were reassuringly similar for all parameters (except the orbital period for obvious reasons), supporting the reliability of the error estimates. The fractional radii are the most important parameters from this analysis and the largest uncertainty estimate for them came from the fitting of subsections of the light curve. They are determined to approximately 1\% precision, which is a good result for a partially-eclipsing system with an eccentric orbit and third light.


\begin{figure}[t] \centering \includegraphics[width=\textwidth]{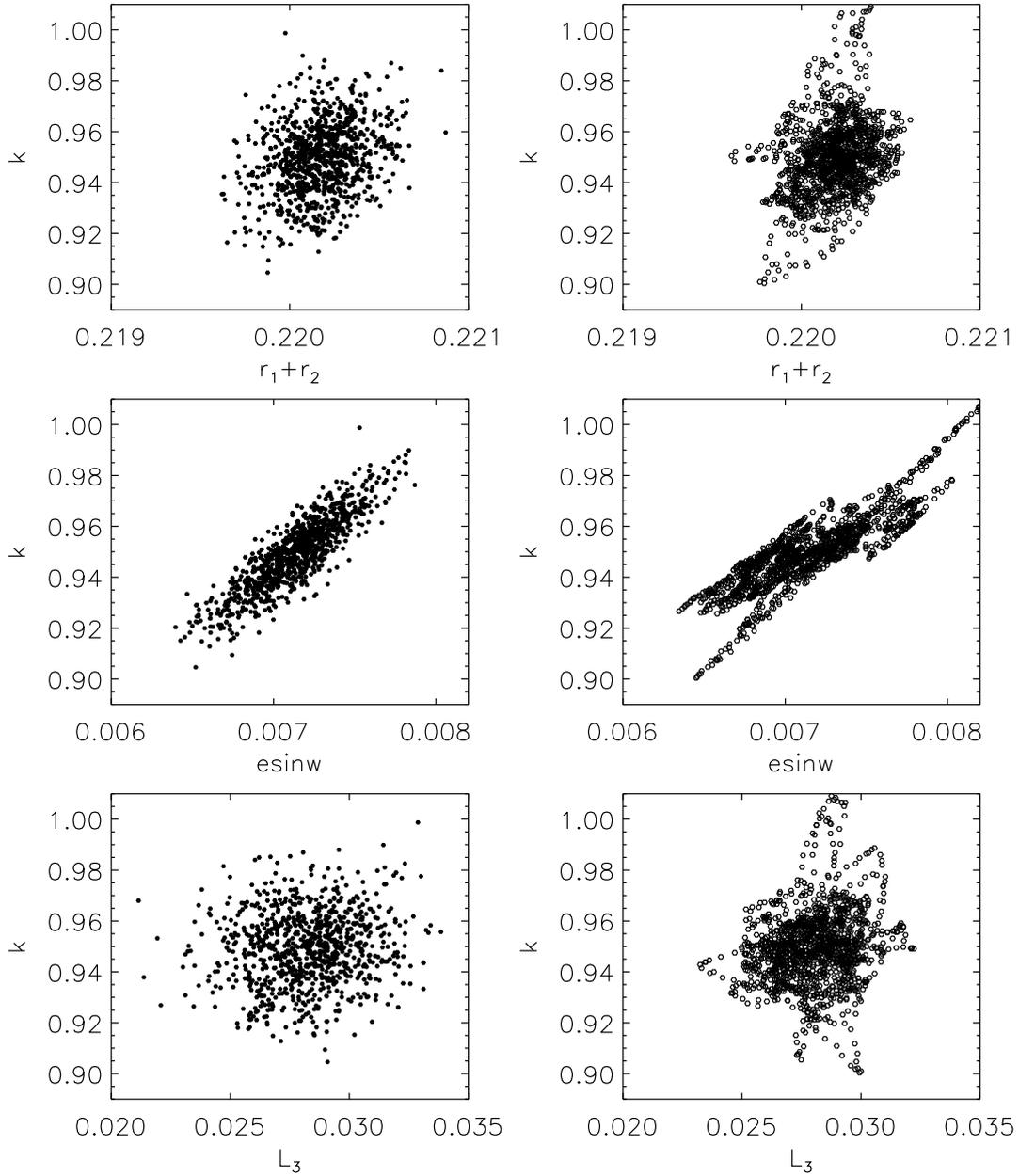} \\
\caption{\label{fig:corr1} Correlation plots for selected photometric parameters.
The panels on the left show results from the Monte Carlo analysis, and on the right
from the residual-permutation analysis. Each point corresponds to the best fit of
one of the synthetic datasets generated during the error analyses. For clarity, 
only the first 1000 points are shown in each case.} \end{figure}

We find an extremely good agreement with the results from the RV observations by Griffin \cite{Griffin13obs} for those parameters in common. They comprise the orbital eccentricity ($0.0099 \pm 00012$ from Griffin versus $0.0099 \pm 0.0003$ here), argument of periastron ($132 \pm 7\degr$ versus $133.7 \pm 1.5\degr$), and the period ($3.1457741 \pm 0.0000012$\,d versus $3.145777 \pm 0.000004$\,d). We also detect the third light to high significance, with a value approximately 13 times its uncertainty, which is the first detection of light from the tertiary component of the system.

To further illustrate the situation we have plotted the results from the Monte Carlo (left panels) and residual-permutation (right panels) algorithms in Fig.\,\ref{fig:corr1}. An immediate conclusion is that the Monte Carlo algorithm appears to be better-behaved, although this is illusory. The paths traced by various solutions in the residual-permutation panels are a result of correlated noise moving gradually through the light curve, meaning successive residual-permutation solutions are closely related. This does not happen for the Monte Carlo solutions, leading to the smoother appearance of the correlation plots from this algorithm.

Fig.\,\ref{fig:corr1} also shows that the sum and ratio of the radii are almost uncorrelated (which is the reason why the fractional radii are parameterised as such in {\sc jktebop}) but there is a clear correlation between $k$ and $e\sin\omega$. Third light is surprisingly not correlated with the ratio of the radii, but in the case of V455\,Aur it is correlated with the orbital inclination and the sum of the fractional radii (not shown).

\begin{figure}[t] \centering \includegraphics[width=\textwidth]{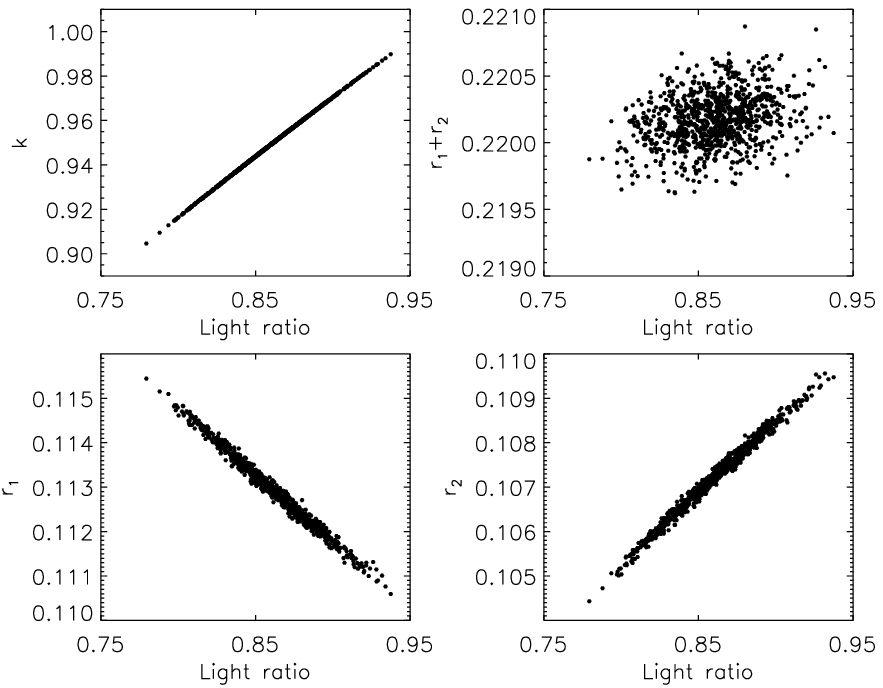} \\
\caption{\label{fig:corr2} Correlation plots for selected photometric parameters. Each point
corresponds to the best fit of a synthetic dataset during the Monte Carlo analysis.} \end{figure}

The primary indeterminacy in the solution is best seen with the ratio of the radii (Fig.\,\ref{fig:corr2}), which is strongly correlated with the light ratio and thus the surface brightness ratio. Because $r_{\rm A}+r_{\rm B}$ is known very precisely from the eclipse durations (its uncertainty in Table\,\ref{tab:lc} is only 0.1\%), $k$ is correlated with $r_{\rm B}$ and anti-correlated with $r_{\rm A}$.

The net result is that the fractional radii of the stars are correlated with the light ratio, and thus a more precise measurement of $r_{\rm B}$ and $r_{\rm A}$ could be obtained if an external constraint on the light ratio of the system were available. We find that if a light ratio with an uncertainty of 5\% were available, the precision of the measurements of $r_{\rm B}$ and $r_{\rm A}$ would be improved from 0.8\% and 1.0\%, respectively, to 0.25\% for both. Such a constraint could be obtained spectroscopically \cite{Torres+00aj,Me++07aa}, helped by the temperatures of the stars being similar. Another possibility is an interferometric measurement, as achieved for the dEB V1022\,Cas \cite{Lester+19aj,Me21obs2}. From the physical properties of the system and the \gaia\ EDR3 parallax we find that the angular separation of the stars is $0.735 \pm 0.003$\,mas. This is within the limits of optical interferometers such as CHARA, but difficult and thus less likely to be achieved versus a spectroscopic measurement of the light ratio.


\section*{Properties of the third component}

Griffin \cite{Griffin13obs} observed the orbital motion of the tertiary star but not its light. He found that it had to be at least 0.5\Msun\ in order to cause the observed orbital motion, but of spectral type K or later to remain undetectable in the cross-correlation functions. Now we have a direct measurement of its light -- from the third light of $\ell_3 = 0.028 \pm 0.002$ found in the {\sc jktebop} analysis -- it is worth revisiting the topic. This amount of third light is small -- especially considering the relatively red response function of \tess\ -- so a low-mass third body is expected.

We began by estimating a mass for the tertiary component. The mass was used to predict the radius and \Teff\ of the third body using the empirical polynomial calibrations\footnote{In Southworth \cite{Me09mn} calibrations of mass versus radius and mass versus \Teff\ were obtained from stars of masses 0.21\Msun\ to 1.59\Msun. The first calibration was used but the second calibration was not; it was retained in that publication in case it became useful in future.} presented in equations 2 and 3 of Southworth \cite{Me09mn}. Its surface gravity was calculated from its mass and radius. We then interpolated the {\sc atlas9} synthetic spectra from Castelli \etal\ \cite{Castelli++97aa} to the \Teff\ and gravity values of the star.

The \Teff\ values of the eclipsing stars were initially taken to be 6340\,K and 6240\,K based on the spectral types advanced by Griffin \cite{Griffin13obs} and the tables of Pecaut \& Mamajek \cite{PecautMamajek13apjs}. These values were subsequently revised to 6500\,K and 6400\,K in a second iteration. Synthetic {\sc atlas9} spectra were interpolated as for the third star, using their surface gravities from Table\,\ref{tab:absdim}. The three sets of synthetic fluxes were converted to relative light contributions using the radii of the three stars, and then passed through the response function of the \tess\ instrument \cite{Ricker+15jatis}. The third light value was calculated as the fraction of the combined light of the three stars that is emitted by the third star. 

We then adjusted the estimated mass of the tertiary component until the predicted amount of third light matched the $\ell_3$ measured from the light curve. In the following section we use the properties of the third star in the process of determining the \Teff\ values of the two eclipsing stars. Once this was completed, we reran the analysis above with the new \Teff\ values in order to arrive at consistent values for the properties of the tertiary component.

\begin{figure}[t] \centering \includegraphics[width=\textwidth]{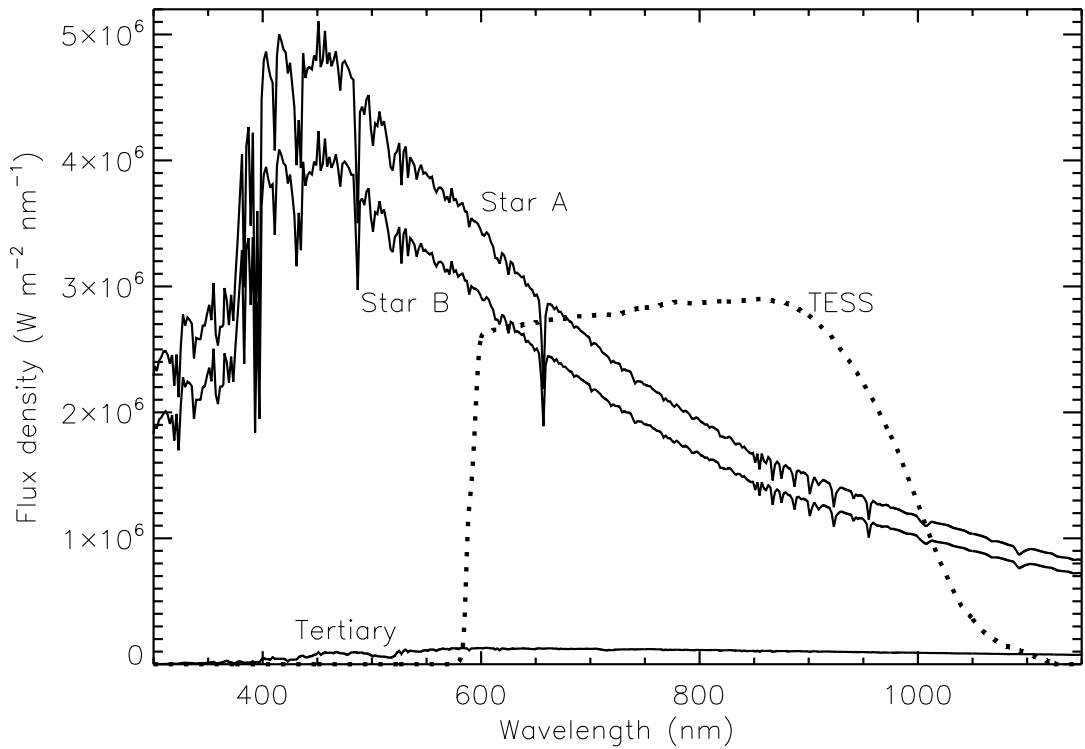} \\
\caption{\label{fig:fluxes} The synthetic spectra of the two eclipsing components
(labelled Star~A and Star~B) and of the third body (labelled Tertiary) from 
{\sc atlas9} model atmospheres and adjusted for the radii of the stars. The 
passband response function from \tess\ is also shown, using dotted lines, and 
has been arbitrarily scaled to half the full height of the plot.} \end{figure}

We found that the tertiary star has a mass of $0.72 \pm 0.05$\Msun, a radius of $0.74 \pm 0.05$\Rsun\ and a \Teff\ of $4300 \pm 300$\,K, where the errorbars include contributions from the third light and also the scatter in the empirical calibrations. The numbers quoted here are those obtained after the iteration to the final \Teff\ values of the two eclipsing stars (see below). A representative plot of the theoretical spectra and the \tess\ passband is shown in Fig.\,\ref{fig:fluxes}. With this mass the inclination of the third orbit is $53 \pm 3\degr$ which implies this triple system is not coplanar.


Using Kepler's third law we find the semimajor axis of the outer (relative) orbit to be 7.6\,au and thus the maximum separation between the eclipsing and the tertiary components is $r_{\rm max} = a(1+e) = 13.0$\,au. At a distance of 78\,pc (see below) this corresponds to an angular separation of 0.17\as. This is resolvable using high-resolution imaging or integral-field spectroscopy of the type often wielded in the detection of close binaries and young planets \cite{Evans++16apj,Beuzit+19aa,Macintosh+14pnas}. Obtaining the spectrum of the third star, or its apparent magnitudes in multiple passbands, would provide additional constraints on its \Teff\ and thus mass. This will be easier at infrared wavelengths due to the brightness of the third component relative to stars A and B: we find that the value of $\ell_3$ is only 0.015 in the $V$-band but increases to 0.082 in the $K_s$-band.


\section*{Physical properties of the eclipsing system}

We used the {\sc jktabsdim} code \cite{Me++05aa} to calculate the physical properties of the eclipsing system and its two components. As input to this we used the fractional radii, orbital period, inclination and eccentricity from the photometric analysis above. To this we added the velocity amplitudes measured by Griffin \cite{Griffin13obs}, $K_{\rm A} = 96.27 \pm 0.16$\kms\ and $K_{\rm B} = 100.73 \pm 0.23$\kms, on the grounds that there was no clear reason to repeat his analysis rather than just adopt his results. We confirmed that the RVs were correctly attributed to the two stars, as it is possible that the photometric primary star (the star obscured at the deeper eclipse) is not the spectroscopic primary star (the star which is brighter or has stronger spectral lines); for examples see Paper IV of this series \cite{Me21obs3} and Theme{\ss}l \etal\ \cite{Themessl+18mn}. The masses and radii of the stars obtained in this way are measured to precisions of 0.5\% and 1.0\%, respectively, which can be attributed to the quality of the RVs from Griffin \cite{Griffin13obs} and the light curve from \tess.

\begin{table} \centering
\caption{\em Physical properties of V455\,Aur defined using the nominal 
solar units given by IAU 2015 Resolution B3 (Ref.\ \cite{Prsa+16aj}). \label{tab:absdim}}
\begin{tabular}{lr@{\,$\pm$\,}lr@{\,$\pm$\,}l}
{\em Parameter}        & \multicolumn{2}{c}{\em Star A} & \multicolumn{2}{c}{\em Star B} \\[3pt]
Mass ratio                                  & \multicolumn{4}{c}{$0.9557 \pm 0.0027$}    \\
Semimajor axis of relative orbit (\Rsunnom) & \multicolumn{4}{c}{$12.295 \pm 0.017$}     \\
Mass (\Msunnom)                             &  1.2887 & 0.0063      &  1.2316 & 0.0050   \\
Radius (\Rsunnom)                           &   1.389 & 0.011       &   1.318 & 0.014    \\
Surface gravity ($\log$[cgs])               &  4.2626 & 0.0070      &  4.2887 & 0.0089   \\
Density ($\!$\rhosun)                       &   0.480 & 0.012       &   0.538 & 0.017    \\
Synchronous rotational velocity (\kms)      &   22.38 & 0.18        &   21.20 & 0.22     \\
Effective temperature (K)                   &    6500 & 200         &    6400 & 200      \\
Luminosity $\log(L/\Lsunnom)$               &   0.492 & 0.054       &   0.419 & 0.055    \\
$M_{\rm bol}$ (mag)                         &    3.51 & 0.13        &    3.69 & 0.14     \\
\end{tabular}
\end{table}

The remaining quantities to be determined are the \Teff\ values of the three stars detectable in the light curve. This can be done by matching the distance from the \gaia\ EDR3 \cite{Gaia20aa} parallax, $77.68 \pm 0.31$\,pc, to that obtained from the radii, \Teff\ values and bolometric corrections of the stars. For this we used the bolometric corrections in multiple passbands from Girardi \etal\ \cite{Girardi+02aa}. We accounted for interstellar reddening of the system by including an estimate of $E(B-V) = 0.002 \pm 0.002$\,mag obtained using the {\sc stilism} \footnote{\texttt{https://stilism.obspm.fr}} online tool (Lallement \etal\ \cite{Lallement+14aa,Lallement+18aa}).

The $BV$ and $JHK$ apparent magnitudes of V455\,Aur, given in Table\,\ref{tab:info}, include the light from all three components but we have the relative light contributions in only the \tess\ passband. We therefore used estimates of the \Teff\ values of the three components and propagated their relative light contributions in the \tess\ band to the other bands \cite{Me+10mn,Me+20aa} using {\sc atlas9} synthetic spectra \cite{Castelli++97aa}. The \Teff\ ratio of the two eclipsing stars is tightly constrained by the surface brightness ratio from the {\sc jktebop} analysis. This process was iterated once in order to ensure consistent properties for all three stars (see previous section). Additional iterations were not needed because the third body contributes only a small amount of light to the system. Once these relative light contributions were determined, we calculated the apparent magnitudes and distance of the eclipsing system. By requiring the distances to be consistent in different passbands and with that from the \gaia\ EDR3 parallax, we determined the \Teff\ values of the eclipsing stars by iterative manual adjustment.

We find the \Teff\ values of the two stars to be $T_{\rm eff,A} = 6500 \pm 200$\,K and $T_{\rm eff,B} = 6400 \pm 200$\,K. These correspond to spectral types of F5\,V and F6\,V, respectively \cite{PecautMamajek13apjs}, which is closer to the F6\,V + F7\,V favoured by Griffin \cite{Griffin13obs} than the F2 in the \textit{Henry Draper Catalogue} \cite{CannonPickering18anhar2}. A detailed spectroscopic analysis to determine precise \Teff\ values and chemical abundances would be very beneficial in improving our understanding of this system \cite{Pavlovski+14mn,Torres+14apj}.

\section*{Summary}

The V455\,Aur system comprises a double-lined spectroscopic binary with a lower-mass third star on a wider orbit. The inner binary was found to be eclipsing from its \hip\ light curve. Extensive RV measurements by Griffin \cite{Griffin01obs,Griffin13obs} yielded precise measurements of the masses of the two eclipsing stars and the first determination of the period of the outer orbit. Prior to the current analysis, however, no photometric study of the system was available.

In this work we have studied the high-quality light curve of V455\,Aur obtained by the \tess\ satellite, which covers seven orbital periods of the inner eclipsing system. We determined precise fractional radii for the two stars and also detected the light contribution of the third star to high confidence. Combining our photometric results with the spectroscopic quantities from Griffin \cite{Griffin13obs}, we have determined the masses and radii of the eclipsing stars to precisions of 0.5\% and 1.0\%, respectively. We have shown that the radius measurements could be further improved by obtaining a spectroscopic light ratio.

The \Teff\ values of the dEB were measured by forcing the distance to the system (determined from the stellar radii and \Teff\ values) to match that found from the \gaia\ EDR3 parallax. As part of this process we subtracted the contribution of the third star from the apparent magnitudes of the system in the $BV$ and $JHK$ bands, and accounted for interstellar extinction using reddening maps \cite{Lallement+14aa,Lallement+18aa}. The mass, radius and \Teff\ of the tertiary was obtained from empirical calibrations based on dEBs \cite{Me09mn} adjusted to match its light contribution to the \tess\ light curve. Its mass is significantly above the minimum mass from the outer spectroscopic orbit, suggesting that the two orbits are not coplanar.

We have made a brief comparison of the properties of V455\,Aur\,AB to the predictions of the PARSEC theoretical stellar models \cite{Bressan+12mn}. We find that the masses, radii and \Teff\ values are reasonably well matched for a fractional metal abundance of $Z = 0.017 \pm 0.003$ and an age of $1.8 \pm 0.2$\,Gyr. The model predictions are slightly too shallow to match the observed mass--radius relation of the two stars, with an agreement at the level of roughly 2$\sigma$. This should be investigated in more detail in future, once futher constraints have been placed on the physical properties of V455\,Aur.

With these new results V455\,Aur can be added to the DEBCat catalogue \cite{Me15debcat}. A detailed spectroscopic analysis of the system is advocated, in order to provide better \Teff\ measurements of the stars as well as a light ratio (helpful in measuring their radii) and photospheric chemical abundances (useful for a comparison with theoretical models). The complication caused by the presence of the third body means this is not a benchmark binary system, as the third light affects the photometric analysis, but the non-coplanarity of the two orbits is interesting from a dynamical viewpoint \cite{Tokovinin18apjs}.


\section*{Acknowledgements}

The current work presented a detailed analysis of an eclipsing binary system that was only made possible by the extensive and impressive RV dataset obtained by Roger Griffin. This work was begun shortly before Roger passed away at the age of 85. I wish to dedicate this paper to Roger, who provided an extraordinary example of what can be achieved via careful analysis of many objects as part of a long-term project. Whilst I never met Roger, I have had the privilege of several lengthy email exchanges with him as well as a brief tour of ``his'' 36-inch telescope at Cambridge. 

I thank Kresimir Pavlovski and Dariusz Graczyk for comments on a draft of this manuscript.
This paper includes data collected by the \tess\ mission. Funding for the \tess\ mission is provided by the NASA's Science Mission Directorate.
The following resources were used in the course of this work: the NASA Astrophysics Data System; the SIMBAD database operated at CDS, Strasbourg, France; and the ar$\chi$iv scientific paper preprint service operated by Cornell University.



\end{document}